\documentstyle{article}
\begin{document}
\centerline{\large On the Physics of Primordial Magnetic Fields} 

\vskip 2cm

\centerline{\bf E. Battaner (1) \& H. Lesch (2)}

(1) Departamento de F\'{\i}sica Te\'orica y del Cosmos. Universidad de
    Granada. Avda. Fuentenueva 18071. Granada. Spain.

(2) Universit$\ddot{a}$ts-Sternwarte M$\ddot{u}$nchen. Scheinerstr. 1,
    D-81679 M$\ddot{u}$nchen, Germany.

\begin{abstract}
There are at present more then 30 theories about the origin of cosmic magnetic fields at galactic and intergalactic scales. Most of them rely on concepts of elementary particle physics, like phase transitions in the early Universe, string theory and processes during the inflationary epoch. Here we present some more
astrophysical
arguments to provide some guidance through this large number and variety of models. Especially the fact that the evolution of magnetic fields depends on the spatial coherence scale of the fields leds to some interesting conclusions, which may rule out the majority of the theoretical scenarios. In principle one has to distinguish between the large-scale and small-scale magnetic fields.
Large scale fields are defined as those as becoming sub-horizon at
that redshift at which the mass energy density becomes equal to the photon energy density, which we name as equality.
Small scale fields which are sub-horizon even before equality, i.e. with scales lower than (present) few Mpc cannot survive the radiation era and cannot reach recombination, because of the effects of magnetic diffusion and photon diffusion. Therefore mechanisms based on phase transitions become unlike, as they provide magnetic fields on scales smaller than the horizon. Thus, the observed galactic and intergalactic fields, which are small scales in our terminology must be created after recombination by normal plasma processes during the protogalactic evolution. The large
scale fields instead were produced during inflation and may have noticeable implications for the formation of the large scale structure of the Universe. The inclusion of large scale magnetic fields may improve present Cold Dark Matter theories of structure formation. \end{abstract}
\section{Introduction}
Primordial and protogalactic magnetic fields have been introduced because of two main reasons: first, they can provide the seed fields for magnetohydrodynamic amplification mechanisms like galactic dynamos; and second, large scale fields could play a role in the formation of structures.

We have no observational knowledge about the first cosmical magnetic fields. Only indirect evidence is available. The magnetic fields in high redshift (2-3) objects like damped Lyman$\alpha$ systems is as strong as in nearby galaxies and galaxy clusters. Since they cannot have appeared immediately with that strength of several $\mu$Gauss they must have started at a considerably weaker strength. The central issue here is whether the magnetic fields are generated by intrinsic plasma processes in the protogalactic fluctuations summarized as {\it battery effects} or whether they are produced deep in the very early Universe by some symmetry breaking during phase transitions or even during the early epochs of inflation just after the Planck time. Battery effects use the different mass of electrons and protons and different collision frequencies of both, with neutral gas or photons. Thereby electric currents are driven, which induce magnetic fields at the first place. Further amplification is provided by magnetohydrodynamical processes, summarized as action of a dynamo. The initial values of the battery fields is of the order of $10^{-20}$ G and can be increased by protogalactic collapse dynamics up to $10^{-14}$G (Chiba and Lesch 1995).

Direct observations are at present restricted to typical redshifts of 3-4 at which the first quasars appear which permits the observation of the widespread intergalactic medium between them and the observer. This renders the subject as speculative as any other cosmological topic. Extragalactic magnetic fields are considered in the review by Kronberg (1994). Readers interested in observations of extragalactic fields are addressed to this review. These observations will be not considered in our contribution which is in part complementary to Kronberg's. Other previous reviews dealing with this topic have
been written by Rees (1987), Coles (1991), Enqvist (1997) and Olesen (1997). Closely
related topics have been considered by Zweibel and Heiles (1997) (Magnetic fields in galaxies and beyond) and Lesch and Chiba (1997) (On the origin and evolution of galactic magnetic fields).

Theoretical models of the origin of magnetic fields were developed in order to satisfy the requirements of astrophysical observations and theories, which have recently been subject to alternative interpretations and improvements. An astrophysical analysis could restrict the large number of different theoretical models, what is the main goal of this review.

Models were developed having in mind that they should provide the seed required by the galactic dynamo, but the efficiency of this dynamo is now
doubtful since the back reaction of the generated magnetic field onto the amplifying plasma flows sets tough constraints onto the maximal achievable field strengths, which are far off the values of several $\mu$G observed in high redshift objects and nearby galaxies (Kurlsrud and Anderson, 1992; Vainshtein and Cattaneo, 1992; Vainshtein, Parker and Rosner, 1993; Cattaneo, 1994; Kurlsrud et al., 1997; Lanzetta, Wolfe and Turmshek, 1995;
Wolfe, Lanzetta and Oren, 1992; Kronberg, Perry and Zukowski, 1992; Perry, Watson and Kronberg, 1993; Kronberg, 1994).

Recent developments (Kulsrud et al. 1997, Lesch and Chiba 1995 Howard and Kulsrud 1997)
indicate that protogalactic magnetic fields are created without any pregalactic seed fields by internal plasma mechanisms, after recombination. Therefore, the value for the pregalactic seed field required by the presence of today galactic magnetic fields may be as low as zero. This fact, together with a better knowledge of the big difficulties for small scale magnetic fields to be conserved and maintained along the radiation dominated era (see Section 3) could render many of the proposed magnetogenesis processes as interesting theoretical exercises without connection with the observable universe. The problem of large scale magnetic fields affecting the formation of large structures could therefore be unconnected with the origin of galactic magnetic fields. It is therefore necessary to precise as much as possible what the astrophysical requirements are at present. 

In absence of loss and production or amplification mechanisms, the frozen-in condition of magnetic field lines would tell us: $$
\vec{B}_0=\vec{B} a^2
$$
being $\vec{B}_0$ the present field and $\vec{B}$ the field when the cosmic scale factor was $a$, taking $a_{0}=1$. As shown by Battaner, Florido and Jimenez-Vicente (1997) this expression is more general, and holds even with no conductivity, under the condition of small perturbations on the Robertson-Walker metrics due to magnetic fields. A pure U(1) gauge theory with the standard Lagrangian is conformally invariant (not like a minimally coupled field), from which it
follows that $\vec{B}$ always decreases following this equation even
in absence of charge carriers (Turner and Widrow, 1988). This equation is of course not true along the whole
evolution of the Universe, because generation, amplification and diffusive losses of the magnetic field became important at some epoch. We will however use equation (1) as a re-definition of $\vec{B_0}$, which will be
therefore no longer the present magnetic field, and will no longer be a constant. This definition is justified because $\vec{B}$ is so much affected by expansion, that the use of $\vec{B_0}$ instead of $\vec{B}$, facilitates the comparison of fields in different epochs. We will call $\vec{B}_0$ the equivalent-to-present magnetic field strength. By adopting (1) we therefore do not pressume any conformally invariance nor any frozen-in condition, but just adopt a definition for $\vec{B}_0$.

Along the paper we will distinguish between large, intermediate and small scales. To be precise we will consider a critical scale $\lambda_{cr}$ defined by: $$
\lambda_{cr}={{1\over{mn_0}}\sqrt{{3 \sigma T^{4}_{0}}\over{8 \pi c G}}}$$ 
where $m$ is the baryon mass, $n_0$ its present number density, $\sigma$ the
Stephan-Boltzmann constant, $T_0$ the present Cosmic Micorowave Background (CMB) temperature, c the speed of light and $G$ the gravitation constant. This length is equivalent to few
Mpc. The criterium is based on the result by Florido and Battaner (1997) who found a very different behaviour for $\lambda<\lambda_{cr}$ and for
$\lambda>\lambda_{cr}$. Physically, $\lambda_{cr}$ corresponds to the size of an
inhomogeneity becoming sub-horizon between Equality and Recombination. It is clear that this transition is very important for our purposes as large scale fields will not be inluenced by any microphysical effect along the radiation dominated era before recombination.

\section{Origin}
The different hypotheses investigating the generation of pregalactic magnetic fields can be classified into four classes, following the epoch of formation: a) during inflation, b) in a phase transition after inflation, c) during the radiation dominated era, and d) after recombination.
\subsection{Magnetic fields generated during inflation} In general, magnetic fields are observed at all scales in the Universe, starting from smallest scales in the solar system,
local interstellar medium up to intracluster scales of several Mpc (Kim et al....) Even if magnetic field inhomogeneities, or coherence cells, have not yet been observed at those large scales as the density structures and CMB anisotropies exhibit, it is natural to expect that magnetic fields these scales exist. As for the case of matter inhomogeneities and radiation anisotropies, inflation provides the most natural explanation of field inhomogeneity, as inflation permits causal connection between two points with a distance that was rather recently, at Equality or slightly later, smaller than the horizon.

Turner and Widrow (1988) first proposed an inflation scenario for the creation of primordial magnetic fields, showing its advantages and difficulties. A cloud with present size $\lambda$ has had at any epoch a size $a \lambda$. This must be compared with the horizon at that epoch, which is a function of $a$. During the first phase of inflation it is rather independent of $a$, becomes $\propto{a^{3/2}}$ during reheating, $\propto{a^2}$ during the radiation dominated era, and $\propto{a^{3/2}}$ during the matter dominated era. Therefore, an inhomogeneity could be sub-horizon when it is produced, becomes super-horizon at a time within inflation and again becomes sub-horizon much later, at Equality, for instance. 

These very-long-wavelength effects were then created by any physical process acting on scales less than the horizon, in practice less than the Hubble radius $H^{-1}$. For a long period in comsic evolution they remained unaffected by local effects and emerged into the causal domain, as the only witness of very early events. Present physical processes may distort the original distribution but the information is not completely lost.

Inflation even provides the excitation mechanism of relative large
wavelength electromagnetic waves out of quantum-mechanical
fluctuations. When these waves reach $\lambda > H^{-1}$, the
oscillating electric and magnetic fields partially appear as static
fields. This is an elegant interpretation of the generation of $\vec{E}$ and $\vec{B}$ in the theory of Turner and Widrow (1988). These fields remain independent of any plasma effect. The conductivity becomes high enough at some time during reheating and will eventually control the small scale magnetic fields. 

As $a(t)$ is exponential during inflation, the initial sub-horizon
scales becomes very large, increasing by a factor greater than
$10^{21}$, which is very suitable for explaining the large structures
and solving the famous horizon problem (e.g. Boerner, 1988). The exponential increase of $a$, presents the main difficulty of classical inflation models for the origin of magnetic fields, because the field decreases
very quickly, following $Ba^{2}=constant$, which is also valid along inflation, irrespective of plasma effects, if the U(1) gauge theory is conformally invariant. Under this invariance, the magnetic strength is reduced by factors of about $10^{104}$. 

Some mechanisms must be assumed to avoid the exponential dilution of magnetic fields. Among other possibilities proposed or analyzed, Turner and Widrow (1988) studied in detail that conformal invariance of electromagnetism is broken through gravitational coupling of the photon. This coupling gives the photon a mass, of the order of $10^{-33}eV$, therefore undetectable. Very interestingly, they were able to predict $B_{0}\approx 5\times 10^{-10}$ at scales of about $1 Mpc$. 

Ratra (1992) considered the coupling of the scalar field responsible for inflation (inflaton) and the Maxwell field, obtaining $B_0$ even as large as $10^{-9}G$ at scales of $H^{-1}/1000$, of about $5 Mpc$, which is also a very promising result, even if the hypothesis is unrealistic in the context of string theory (Lemoine and Lemoine, 1992). Garretson, Field and Carrol (1992) invoked a pseudo-Goldstone-boson coupled to electromagnetism, obtaining very low values ($B_{0}<10^{-21}G$ at $\lambda=1 Mpc$). Dolgov (1993) proposed the breaking of conformal invariance through the so called "phase anomaly", a mechanism that would not work in the supersymmetric theory (Gasperini, Giovannini and Veneziano, 1995b; Lemoine and Lemoine, 1995). Dolgov and Silk (1993) considered a spontaneous break of the gauge symmetry of electromagnetism that produced electrical currents with non-vanishing curl.

The model by Davis and Dimopoulos (1995) is based on the creation of magnetic fields at the GUT phase transition (and therefore has much in common with models commented in the next section, but this transition could take place during the inflation period). They predicted values as high as $10^{-11}G$ at galactic scales. 

Considering the earlier Planck-scale Universe could help in discriminating this profusion of theoretical viable models of inflationary magnetogenesis (Lemoine and Lemoine, 1995). In the inflactionary "pre-big-bang" scenario based on the superstring theory (Veneziano, 1991; Gasperini and Veneziano, 1993a,b, 1994) the electromagnetic field is coupled not only to the metric but also to the dilaton background. COBE anisotropies emerge from electromagnetic vacuum fluctuations (Gasperini, Giovannini and Veneziano, 1995a), involving scales of the order of $100 Mpc$. For some values of arbitrary parameters, these models provide large enough values of (inter)galactic magnetic fields, even in the absence of galactic dynamos (Gasperini, Giovannini and Veneziano, 1995b). They are in fact able to explain a possible equipartition of energy between the CMB radiation and magnetic fields. (See however section 4, where it is argued that this equipartition, if real, has been reached later). Hence, the "pre-big-bang" scenario is able to provide strengths and scales as required by present astrophysical observations. 

\subsection{Magnetic fields generated in phase transitions} Hogan
(1983) gave the basic arguments to consider phase transitions of first
order as potential mechanisms for the generation of primordial
magnetic fields. The phase transition would not take place
simoultaneously in all places of the Universe, but in causal
bubbles. At the rim of the bubbles very high gradients of the
temperature, or any other order quantity characterizing the phase
transition, such as the Higgs vacuum expectation value, would be stablished. These high gradients
would produce a thermoelectric mechanism akin to the Biermann battery (Biermann, 1950; Biermann and Schlueter, 1951; Kemp, 1982). When bubbles collide the fields from each bubble are stitched to those of their neighbors by magnetic reconnection and the magnetic field lines execute a Brownian walk, related to the future spectrum of magnetic fields.

The electroweak phase transition has been considered by Vachaspati (1991), Enqvist and Olesen (1993, 1994), Davidson (1995), Grasso and Riotto (1997), Tornkvist (1997), Hindmarsh and Everett
(1997) and others. The QCD phase transition has been considered by Quashnock, Loeb and Spergel (1989), Cheng and Olinto (1994), Sigl, Olinto and Jedamzik (1996) and others. The GUT phase transition has been considered by Brandenberger et al. (1992), Enqvist and Olesen (1994), Davis and Dimopoulos (1995), Martins and Shellard (1997) and others. There is a large variety of points of views and treatments other than the original of Hogan less less less less lesn (1983) and Vachaspati (1991). Vachaspati and Vilenkin (1991) considered cosmic strings formed in phase transitions with wiggly motions which created vorticity and then magnetic fields. Baym, Boedeker and McLerran (1996) showed how second order phase transitions also can generate magnetic fields. Kibble and Vilenkin (1995) considered the possibility that magnetic fields could also be generated in the intersecting region of two colliding bubbles. See also the review by Enqvist (1977). 

It is interesting to note that the predicted present spectrum of
magnetic field inhomogeneities is rather independent of the nature and
time of the phase transition. Suppose that $\lambda_i$ is the
correlation length at the phase transition taking place at a
temperature $T_i$. We have $B_i$ (the magnetic field produced at the
phase transition) to be of the order of $T^2_i$ and
$\lambda_i=T^{-1}_i$ (Vachaspati, 1991). The present magnetic field
corresponding to this scale $T^{-1}_i z_i$ in comoving coordinates,
$B_{oi}=B_i R^2_i=T^2_i R^2_i=T^2_0$, being $T_0$ the present CMB
temperature. It is independent of subindex $i$, characterizing the
phase transition. (We have used units taking $c=h=k=1$). To calculate
the spectrum we have $B_0(\lambda)\approx B_{0i}/N$ (Vachaspati, 1991)
where $N=T_0 \lambda$ is the number of correlation cells at the scale
$\lambda$ of interest, therefore $B_0(\lambda)=T^2_0/(T_0 \lambda)=
T_0/\lambda$, which does not contain subindex $i$. There is a
compensation of two effects: The higher $T_i$, the higher the magnetic field produced, but the larger the effect of dilution by expansion. Other authors propose not to divide by $N$ but by $\sqrt{N}$ (Enqvist and Olesen, 1993). This changes the spectrum but it is still rather independent of the phase transition involved. An important consequence of this fact is that the effect of the different phase transitions could be added to produce an enhanced spectrum.

However, one of the big problems encountered by phase transitions in general, as magnetogenesis mechanisms, is that they provide very small values of the magnetic field at galactic scales. For instance, Vachaspati (1991) found $B_0\approx 10^{-30} G$. This result was improved by Enqvist and Olesen (1993) as they divided by $\sqrt{N}$ and not by $N$, but even so they obtained $B_0\approx 4\times 10^{-19}G$, enough to become the seed for galactic dynamos, but insufficient for the large values in protogalactic objects, or if the galactic dynamos do not work. Quashnock, Loeb and Spergel (1989) found $6\times 10^{-38}$; Vachaspati and Vilenkin, $10^{-27}$; etc. Beck et al. (1996) summarizing previous works, gave a value less than $10^{-23}$.

Related to the above problem, these models provide very small scales, and the magnetic field would be destroyed by microphysical mechanisms in the radiation dominated era. This will be discussed in Section 3. 

\subsection{Magnetic fields generated in the radiation dominated era}
Matsuda, Sato and Takeda (1971) first proposed a turbulent dynamo working in a radiation dominated universe. Even if the existence of a cosmological turbulence was already under discussion at that epoch, the interesting paper by Harrison (1973) considered that magnetic fields were generated by turbulent vorticity. The treatment was not relativistic. The turbulent medium was made up of ions and a negatively charged dense component composed of electrons and photons tightly coupled by Thompson scattering. A close relation between vorticity and the magnetic field was found: $$
\vec{B}=-(mc/e)\vec{\omega}
$$
where $m$ and $e$ are the proton mass and charge and $\vec{\omega}$ is the vorticity. This relation was already observed by Batchelor (1950) and has been recently reconsidered by Kulsrud et al. (1997). Magnetic field strengths of the order of $5\times 10^{-4}G$ and a characteristic scale
of $2\times 10^2$pc were obtained for protogalaxies and $5\times 10^{-8}G$ and $10$
kpc for the intergalactic medium. It is interesting to note that in this model there was an "external" scale in which structures and peculiar motions were frozen, with a "turbulent horizon" being a small fraction of the Hubble radius. The existence of a primordial turbulence is still doubtful (see for instance the review by Rees, 1987). The required primordial vorticity, a critical point in this scenario, has been reconsidered by Sicotte (1997). 

\subsection{Magnetic fields generated after Recombination} 

During recombination the plasma decoupled from the radiation
field. The protons caught the free electrons until the decreasing
temperature and density did not allow for further recombination. A gas
was left which was only partially ionized with $n_i/n_H\sim 10^{-4-5}$
(e.g. Peebles 1993) and in which the density fluctuations should lead
to the formation of galaxies. For our purposes it is enough to state
that during the epoch of galaxy formation enough processes appear
which can explain the existence of magnetic fields via battery
mechanisms, which are a necessary ingredient for magnetohydrodynamical
models for galactic magnetic fields, since in basic hydromagnetic
equation $$ 
{\partial {\bf B}\over {\partial t}}=\nabla\times({\bf v}\times{\bf
B})- {\bf \nabla\times}\eta({\nabla\times\bf B}). 
$$
no source term for the magnetic field appears. That is to say, there is no outright creation of magnetic field in the hydromagnetic description of galactic magnetic fields. Hence, if at any time the universe was devoid of magnetic fields, then as far as hydromagnetic effects are concerned, there would be no magnetic field at any other time. 

The battery mechanisms provide these {\it seed fields}. We describe now how
te seed
fields are related to the protogalactic density fluctuations after their decoupling from background radiation, i.e. after recombination (Lesch and Chiba 1995). Whereas for primordial magnetic fields phase transitions and symmetry breaking mechanisms have to be considered, the generation of magnetic fields after recombination is described in terms of elementary
electrodynamic properties of a plasma consisting of electrons and
protons. The essence of any battery process is that currents are
produced whenever the mean velocities of negative and positive charge
carriers differ. In general, negative charge carriers are electrons
and as such they are orders of magnitude less massive than the
positive charge carriers. This makes electrons more responsive to
inertial drag forces than ions are. The combination of a gravitational
field with differential rotation leads to a nonconservative force
acting essentially upon the electrons. These two ingredients occur
naturally in disk systems with a central radiation source, as well as
in stars, as was first pointed out by Biermann (1950). The ions are
concentrated to the equitorial plane by the generated electric
field. However, this field cannot cancel the centrifugal acceleration
completely, i.e. charges must move and meridional currents
appear. Lesch and Chiba (1995) transferred Biermann's battery into the
context of a forming galaxy. The resulting magnetic field of a
spherical over-dense region in an expanding universe, with an
expansion rate $a = 1/(1+z)$, is governed by the following equation 
$$
{1\over{a^2}}{\partial\over{\partial t}}a^2B={m_i c\over{2e}} |{\bf
\nabla\times g}|\simeq{m_i c\over{2e}}\omega(t)^2. $$

where ${\bf g}$ denotes the centrifugal acceleration, $\omega$ is the rotation velocity, and $m_i$ is the ion mass.

A second battery process was invoked by Mishustin \& Ruzmaikin (1973). They considered the interaction of a rotating electron-proton-plasma with the intense cosmic background radiation. Thermal electrons scatter the photons of the background radiation via Compton scattering and gain energy and momentum, thereby drifting relative to the protons, i.e. producing a current. This current induces a magnetic field, whose time evolution is described by
$$
{1\over{a^2}}{\partial\over{\partial t}} a^2 B ={m_e c\over
e}{2\omega\over{\tau_\gamma}}. 
$$
$\tau_\gamma$ denotes the optical depth for Compton scattering, which is a sensitive function of the redshift
$$
\tau_\gamma ={3 m_e c\over{4\sigma_T\rho_\gamma(0)(1+z)^4}}\equiv \tau_\gamma(0)(1+z)^{-4}.
$$
$\sigma_T=6.65\cdot 10^{-25}$ cm$^2$ is the Thomson cross section and $\rho_\gamma\simeq 4\cdot 10^{-13}$ erg cm$^{-3}$ is the present energy density of the background radiation. Here a term related to the Coulomb collisions of electrons and protons is neglected compared to the effect of current generation in the context concerned (Mishustin \& Ruzmaikin 1973). 

Magnetic fields are also created
by sheared flows in weakly ionized plasmas as proposed by Lesch et
al. (1989) for active galactic central regions, and by Huba \& Fedder
(1993) for the general case of shearing motions between plasmas and
neutral gases: Again, the different mobility of electrons and ions is
used. The electrons collide with neutral atoms, thereby drifting
relative to the ions. For a differentially rotating system, the drift
corresponds to a current, which induces a magnetic field. The field
generation term is 
$$
{1\over{a^2}}{\partial\over{\partial t}} a^2 B ={m_e c\over e}|{\bf
\nabla\times} \nu_{en} ({\bf V_i - V_n})| \simeq{m_e c\over e}\nu_{en}
{v_r\over{l_{shear}}}. 
$$
$\nu_{en}$ denotes the electron-neutral collision frequency, $v_r$ is the relative ion-neutral drift speed, and $l_{shear}$ is the shear length. ${\bf V_i} \ ({\bf V_n})$ is the ion (neutral) fluid velocity. 
The battery effects discussed above result in field strength at the so-called "turnover" of about $10^{-23}-10^{-19}$ G. The turnover denotes the redshift at which a gravitationally unstable structure decouples from the overall Hubble expansion. Taking into account the dynamics of a collapsing disk galaxy Lesch and Chiba (1995) showed that the field strengths at that redshift at which galactic disks form is at least four to five orders of magnitudes stronger then the field at turnover. The battery mechanisms within protogalactic systems lead to seed fields between $10^{-13}-10^{-16}$G. During the disk formation non-axisymmetric instabilities lead to further now exponential growth of the field on time scales of the order of $10^8$ years by compression (Chiba and Lesch 1994). So after about 1-2 Gigayears the forming galaxies will contain $\mu$G-fields, as it is observed in the high-redshift Lyman$\alpha$-clouds

We have seen the possible production and evolution of seed magnetic fields in the course of the growth of protogalactic density fluctuations. In this picture, principal ingredients of presently observed magnetic fields are supposed to be seeded after the recombination epoch {\it and} before the first ignition of stars in a disk. Magnetic fields that came out in a forming disk galaxy are compatible with those reported in high redshift objects.

Alternative scenarios of seeding galactic magnetic fields have also been proposed, invoking the detailed plasma processes in galactic nuclei and jets, the effects of first star formation, and the pregalactic physics. 

Some of extragalactic sources reveal radio jets (e.g. Bridle and Perley 1984). The magnetic fields in jets, typically having equipartition strength of $\sim 10 \mu G$ and coherent length of several tens kpc, are thought to originate in the vicinity of a central compact object. Daly and Loeb (1990) argued that if all of galaxies were initially activated through a compact nucleus, which are currently inactive, the jets associated with a nucleus must tunnel through the ambient protogalactic medium, and the equipartition magnetic fields carried by jets are dispersed over the ambient medium with the strength compatible with the present strength of several $\mu G$. The picture presented is devoted to the fate of observed jet magnetic fields; original magnetic fields near a central compact object may be seeded by inherent plasma process involved (Lesch et al. 1989; Chakrabarti 1991) or accretion from the body of the host galaxy. The magnetic fields near the center may be further
strengthened up to equipartition strength by accretion-disk dynamos around a central object (Bisnovati-Kogan and Ruzmaikin 1976; Pudritz 1981), and the velocity gradient in jets results in the longitudinal field. 

Alternatively, the production of seed fields may have had to await to the onset of star formation; although a process of forming first-generation stars is very different from the present if there were no magnetic fields (Rees 1987), once the stars formed, the combination of battery and dynamo mechanisms inside stars may generate magnetic fields. The fields then could be ejected into the interstellar medium via stellar winds and supernova explosions. The resulting magnetic field is randomly aligned with a characteristic scale of 100 pc and strength of $\mu G$ (e.g. Ruzmaikin et al. 1988).
Supposing that the interstellar medium of damped Ly$\alpha$ clouds is enriched by first generation of stars, it may also be enriched by magnetic fields created within stars if the picture described works. However the field flux is mostly concentrated on a number of small-scale components with many reversals, and thus it is hard to imagine how the ensemble of such small-scale fields reproduces the statistically siginificant level of Faraday rotation. Some sort of pregalactic dynamos may be responsible for organizing the large-scale structure of magnetic fields (Zweibel 1988; Pudritz and Silk 1989).

\section{Evolution of small scale fields} 

After Annihilation and before Recombination, small scale magnetic fields are subject to severe destructive microphysical processes. Magnetic fields are then supported by electric currents that must be stablished or maintained in a very dense photon medium very effectively interacting with electrons and protons through Thomson scattering. Lesch and Birk (1998) have studied the conductivity and hence the magnetic diffusion during this critical epoch. They gave an equation for the diffussion time equivalent to: $$
\tau_{diff}=10^{44} z^{-6} \lambda^2
$$
where $\tau_{diff}$ is measured in seconds and $\lambda$ in cm. The dependence of $\tau_{diff}$ on $z$ is very pronounced, the most restrictive action of the magnetic diffusivity is at the beginning, for $z=z_{ann}$, at Annihilation. The question is to find what scales are able to survive and reach Recombination, when this hostil medium is decoupled. Taking $\tau_{diff}=\tau_{rec}$ (recombination) and $z=z_{ann}$ we obtain:
$$
\lambda=5\times 10^{-16} z^3_{ann}
$$
and this $\lambda$ will grow to its present comoving size: $$
\lambda_0=5\times 10^{-16} z^4_{ann}
$$
The Annihilation took place at $T=5 \times 10^9 K$ (being the mass of the electron $0.511 MeV$), therefore, $z_{ann} \approx 2 \times 10^9$. We conclude that only scales greater than about $3 kpc$ would survive.

This is really a large value if the field was generated by a phase transition. The most recent one, the QCD phase transition, took place at $\sim 200 MeV$, with a correlation scale of $T^{-1}_{QCD}$ (Vachaspati,1991), i.e. $10^{-11} cm$ in conventional units. This is at present only $10 cm$. Other phase transitions provide similar values, as the correlation length at any phase transition corresponds to a present size of $z_{pt}/T_{pt} \sim (T_{pt}/T_0)/T_{pt} \sim t^{-1}_0$, noticeably independent of the precise phase transition involved. The subindex $pt$ denotes any phase transition. 

The minimum value of $\lambda_0 = 3 kpc$ is even much higher than the
present size of the horizon at any phase transition. This is important
as probably the correlation length grows faster than the horizon
(Dimopoulos and Davies, 1996) and then $\lambda_0$ should be compared
with the horizon present size. The horizon at the QCD phase transition was $\sim 10^6 cm$, equivalent to $\sim 0.2 pc$ at present. The horizon at the electroweak phase transition was only few centimeters, corresponding to about $1 AU$ at present. For earlier phase transitions the situation is even worse. 

Some mechanism for very efficiently increasing the scales is required. An interesting calculation was carried out by Brandenburg, Enqvist and Olesen (1996). These authors considered MHD in a turbulent expanding radiation dominated universe, the metric variations being ignored. They found an inverse cascade, producing larger and larger scales for increasing time. The energy of small scale fields is transferred to larger scale fields. Important as it is, this model seems to be insufficient.

The existence of a turbulence, i.e. of
non-linear effects in a medium so extremely close to equilibrium (much more than the CMB) is controversial. Aside the remarks summarized by Rees (1987), the relative contrast density $\delta\equiv\delta\rho/\rho$ evolves not in a random way, as it could be expected from a turbulent behaviour. The Jeans mass is very low, in particular at the beginning of the radiation dominated epoch, of less than $1 M_\odot$ (e.g. Battaner, 1996), so that collapse is a common state of inhomogeneities. If $\delta>0$ initially $\delta$ will always increase, and if $\delta<0$ initially $\delta$ will always decrease, at least for scales equivalent to a baryon rest mass less than $1M_\odot$. Autogravitation, or in the required relativistic treatment for this epoch, perturbations of the metric tensor, is a basic fact in the evolution of inhomogeneities. Gravitational collapses, even if they do not process very fast, only $\propto a^2$, cannot be ignored. Even if it is difficult to concive turbulence in a medium dominated by collapses, perturbations of the metric tensor should be incorporated to this type of turbulence models. Of course, also,
turbulence and inverse cascades must stop at scales comparable to the horizon (Harrison, 1973), therefore being unable to explain fields at comoving scales larger than $1 Mpc$.

Quantitatively, for the smaller
scales, the obtained figures seem to be too low. The larger scale feeded are only of the order of 2 pc and the N value of Vachaspati (1991) is shifted from about $10^{24}$ to $10^{19}$, clearly insufficient. The calculation is limited to a time $10^9$ times the electroweak phase transition time. Extended calculations to much recent times could provide much less values of N, so this model can still deserve interesting possibilities. But we find unlike that this mechanism was able to surmount the effects of a so large conductivity,
if a turbulence actually exist at all, during this epoch. A relativistic MHD in an expanding universe has been also studied by Gailis, Frankel and Dettman (1995).

There is another effect destroying small scale magnetic fields along the radiation dominated era and specially just after Equality and before Recombination. On general grounds, one would expect that magnetic field inhomogeneities should be associated to radiation and matter inhomogeneities and that the former would be destroyed if the later are damped. A classical treatment of density inhomogeneities in the imperfect fluid made up of photons and baryons (Weinberg, 1972; Silk, 1968) shows that masses less than the Silk mass are damped in the Acoustic epoch, when the cloud mass becomes larger than the Jeans mass, before Recombination. It is unlikely that magnetic fields prevent the inhomogeneity from the destructive effects of viscosity and heat conduction due to photon diffusion.

A model of this magnetized imperfect fluid has been developed by Jedamzik, Katalinic and Olinto (1996) concluding that MHD modes are completely damped by photon diffusion up to the Silk mass, as expected, and convert magnetic energy into heat. Damping would also be very important during the neutrino decoupling era, therefore small scale fields could have been washed out before the radiation dominated era. A direct consequence of photon diffusion damping would be that primordial magnetic fields would neither directly produce present galactic fields nor directly influence the galaxy formation process. 

An equivalent argument is given by Lesch and Birk (1998) showing that vorticity and their potentially associated magnetic fields are severely affected by kinematic viscosity.

However, Brandenburg, Enqvist and Olesen (1977) are again more optimistic, estimating that the inverse cascade process is scarcely affected by Silk damping, except very late and perhaps for very weak fields.

\section{Effects of large scale magnetic fields} Large scale magnetic fields are not affected by microphysical processes and evolve as $B \propto a^{-2}$, or in other words $B_0$ is constant, more or less, from Inflation to Recombination. Even after Recombination, the evolution should not be dramatic. Large scale density inhomogeneities still
behave linearly and so probably behave their (probably) associated magnetic field inhomogeneities. Small scale effects such as ejections from radiogalaxies, dynamos, contractions in galaxy formation, non-linear effects etc., taking place at protogalactic stages or once the first galaxies are formed do not alter the $B \propto a^{-2}$ evolution of large scale fields. Shapes of field configuration are conserved, just they grow within expansion, becoming larger and weaker.

These large scale $\vec{B}$-inhomogeneities may have had a substantial influence on the formation of the large scale structures in the Universe. Since long, several authors have considered that magnetic fields could affect the formation of galaxies, mainly Piddington (1969, 1972) who tried to explain the present morphological differences between different types of galaxies from differences of magnetic and angular directions when galaxies formed. Wasserman (1978) proposed that magnetic field configurations at Recombination could decide the formation of galaxies and even their angular momenta. This work has been recently continued and extended to the non-linear regime by Kim, Olinto and Rosner (1996), being this model and the pioneer one by Wasserman (1978) were devoted to the post-Recombination era. It probably happens however, that the magnetic fields and the density structure were formed before, during the radiaton dominated era or before. These models mainly consider the problem of the formation of galaxies, while we are here favoring that this process is not directly affected by primordial magnetic fields.

Let us consider the problem of how large scale magnetic fields inhomogeneities have an influence on the formation of large scale density inhomogeneities in the Universe. Coles (1992) pointed out that the failure of the CDM scenario to explain large scale structures could be satisfactorely surmounted if magnetic felds were taken into account. The observations of large structures is in clear disagreement with the random behaviour predicted by CDM models, showing an impresive regularity and periodicity (Einasto et al., 1997).

The study of the influence of $\vec{B}$ along the large scale structure along the radiation dominated epoch was undertaken by Battaner, Florido and Jimenez-Vicente (1997), Florido and Battaner (1997) and Battaner, Florido and Garcia-Ruiz (1997), introducing linear perturbations in the physical quantities, including the metric tensor and the magnetic field, in a Robertson-Walker metrics. They found that preexisting magnetic structures were able to produce anisotropic density inhomogeneities in the photon fluid and local perturbations of the metrics.
In particular, they were able to produce filaments. These radiative and gravitational
potential filaments were the sites where baryons, or any other dark matter component, collapsed, forming the today observed luminous filaments as elements of the large scale structure (Shectman et al., 1996). Magnetic fields of the order of $B_0=10^{-8}-10^{-9}G$ could be responsible of the filamentary large scale structure. Cosmological filaments, as any other small scale filament in astrophysical systems could be interpreted as a magnetically driven configuration. Araujo and Opher (1997) have also considered the formation of voids by the magnetic pressure. 

If the large scale is made up of filaments joining together to produce a network, and if these filaments actually were magnetic in origin, then the network would be subject to some magnetic restrictions, arising from the condition $\nabla.\vec{B}=0$ and from reconnection processes. Battaner, Florido and Garcia-Ruiz (1997) carried out a cristalographic approach, showing that the simplest network under these conditions was an "egg-carton", formed up by octahedra joining at their vertexes. This "egg-carton" universe would have larger amounts of matter along the edges of the octahedra, and specially at the vertexes which would be the sites of large superclusters of galaxies, and voids would correspond to the interiors of the octahedra. From the nodes of the lattice, were two octahedra join, eight filaments would emerge. This spider-like structure has been observed for the local supercluster (Einasto, 1992). It is otherwise very difficult
to explain the extreme regularity observed (Tully et al., 1992; Einasto et al.,1997).

Magnetic fields should not be considered as an alternative to current theories on large scale structure formation, but rather, as suggested by Coles (1992), as a missing ingredient in them. 

\section{Limits and future observations} Several limits to the magnetic field intensity or energy density have been reported in the literature (see also the reviews by Lesch and Chiba, 1997, and Beck et al., 1996). However, most of these limits affect a cosmological homogeneous magnetic field (which is a hypothesis scarcely defended; esceptions are Zeldovich, 1965, and Enqvist and Olesen,1994) therefore being useless if magnetic fields were randomly distributed (with $<\vec{B}>=0$ even if $<B^2>\neq0$), or at
least if there existed a homogeneous distribution of magnetic energy density, which is
probably also a bad assumption. If instead, we are interested in the limits of typical peak values, the above mentioned limits should be increased by a factor which would depend on the statistical distribution of size and position of coherence cells or filaments. This factor could be of the order of 100 or 1000. This consideration affects, for instance, the limits based on the $^4He$ abundance of about $B_0\le 10^{-7}G$ (Greenstein, 1969; Zeldovich and Novikov, 1975; Matese and O'Conell, 1970; Barrow, 1976; Cheng , Scharmm and Truran, 1994; Kernan, Starkman and Vachaspati, 1995; Grasso and Rubistein, 1995, 1996; Cheng et al., 1996, and others), on the neutrino spin flip of about $B_0\le 4\times 10^{-9}G$ (though very much depending on the mass of all neutrinos) (Shapiro and Wasserman, 1981, Enqvist et al., 1993) and on the CMB isotropy of about $B_0 < 4\times 10^{-9}G$ (Lesch and Chiba, 1997; Barrow,
Ferreira and Silk (1997).

From an observational relation between Faraday rotation and redshift of 
quasars it is concluded a limit for a widespread cosmological aligned field of about $<10^{-11}G$ (Rees and Reinhardt, 1972; Kronberg and Simard-Normandin, 1976; Vallee, 1983; Lesch and Chiba, 1997). This limit is weakened to $10^{-9}G$ if the coherence cells are $1Mpc$ large (Kronberg, 1994) or weakened to become $>3\times 10^{-8}G$ if the field is coherent only on scales $<10Mpc$ (Kosowsky and Loeb, 1996). Peak values in a structure similar to the mentioned in the above section could be much greater than these limits for the same Faraday rotation data.

Let us propose a limit for a typical maximum of $B_0$ in the radiation dominated era. Rees (1987) estimated that in order to trigger galaxy formation, magnetic fields just after recombination would amount to $B_0>10^{-9}G$. The argument could be inverted to provide un upper limit. Based on the results by Battaner, Florido and Jimenez-Vicente (1997) and Florido and Battaner (1997) we must have:
\begin{equation}
B_0<10^{-8}G
\end{equation}
for large scale peaks in the radiation era, because otherwise the formation of large scale structures would have begun too early and would be at present in a much advanced state of collapse.

Clearly, we observe at present peak values much larger than these. Today, if we exclude small scale peaks, such as jets, or even pulsars, we could have $B_0\approx 10^{-6}G$ (Kronberg, 1994). Therefore, some post-Recombination processes have either amplified or generated additional intergalactic fields. 

Observations of present intergalactic and protogalactic magnetic fields have been reviewed by Kronberg (1994) and the results should not been repeated here. Let us therefore comment some recent proposals of future potential observations.

Plaga (1995) has proposed that the arrival time of $\gamma$-rays from extragalactic sources could provide information about very low intergalactic magnetic fields, in the range $10^{-12}-10^{-24}G$. The delay in the arrival of the energetic TeV-$\gamma$-rays, with respect the low energy $\gamma$-rays that would directly reach us, is due to $e^-e^+$ pair production involving IR background radiation. The particle pairs produced would scatter off CMB photons, producing the observable high energy $\gamma$-rays. See also the comment by Kronberg (1995) on this method.

Observations of coherence cells of aligned disc warps (Battaner et al., 1991; Zurita and Battaner, 1997), under the interpretation that these warps are produced by intergalactic magnetic fields (Battaner, Florido and Sanchez-Saavedra, 1990) have provided temptative values of $\lambda\approx25 Mpc$. Future $21 cm$ and optical galactic maps and surveys could provide better results and extended to greater regions in the Milky-Way neighborhood.

Improving the sensitivity of experiments measuring the CMB radiation, in a feasible way in a next future, would also permit to gather information about magnetic fields (Magueijo, 1994). Kosowsky and Loeb (1996) analized their influence on the Faraday rotation of the CMB radiation estimating that a field of $10^{-9}G$ would produce a Faraday rotation of 1 degree at a frequency of $30 GHz$. Adams et al. (1996) proposed that $10^{-9}G$ fields generated at inflation would produce measurable distortions in the acoustic peaks in the CMB radiation.

Observations of the composition, spectrum and directional distributions of extragalactic ultrahigh energy cosmic rays, with energies greater than $10^{18}-10^{19} eV$ can deserve estimates on the large scale component of magnetic fields of the order of $10^{-9}G$ or less (Lee, Olinto and Sigl, 1996; Stanev et al., 1995).

\newpage

J. Adams, U.H. Danielson, D. Grasso and H.R. Rubinstein,
Phys. Lett. B {\bf 388}, 253 (1996).

J.C.N. de Araujo and R. Opher, astro-ph/9707303 (1997).

J.D. Barrow, Mon.Not. R.A.S. {\bf 175}, 339 (1976).

J.D. Barrow, P.G. Ferreira and J. Silk, astro-ph/9701063 (1997).

E. Battaner, {\it Astrophysical fluid dynamics}, Cambridge
Univ. Press (1996).

E. Battaner, E. Florido and J.M. Garcia-Ruiz,
Astron. Astrophys. in press (1997).

E. Battaner, E. Florido and J. Jimenez-Vicente,
Astron. Astrophys. in press (1997).

E. Battaner, E. Florido and M.L. Sanchez-Saavedra,
Astron. Astrophys. {\bf236}, 1 (1990).

E. Battaner, J.L. Garrido, M.L. Sanchez-Saavedra and E. Florido,
Astron. Astrophys. {\bf 251}, 402 (1991).

G.K. Batchelor, Proc. R. Soc. London {\bf å201}, 405 (1950).

G. Baym, D. Boedeker and L. McLerran, Phys. Rev. D {\bf 53}, 662
(1996).

R. Beck, A. Brandenburg, D. Moss, A. Shukurov and D. Sokoloff,
An. Rev. Astron. Astrophys. {\bf 34}, 155 (1996).

L. Biermann, Zeit. Naturforschung {\bf 5a}, 65 (1950).

L. Biermann and A. Schleuter, Phys. Rev. {\bf 82}, 863 (1951).

G.S. Bisnovatyi-Kogan and A.A. Ruzmaikin, Astrophys. Space
Sci. {\bf 42}, 401.

G. Boerner, {\it The Early Universe}, Springer-Verlag. Berlin
(1988).

R.H. Brandenberger, A.C. Davis, A.M. Matheson and M. Thodden,
Phys. Lett. B {\bf 293}, 287 (1992).

A. Brandenburg, K. Enqvist and K. Olesen, Phys. Lett. B {\bf
392}, 395 (1997).

A. Brandenburg, K. Enqvist and P. Olesen, Phys. Rev. D {\bf 54},
1291 (1996).

A.H. Bridle and R.A. Perley,  Ann. Rev. Astron. Astrophys {\bf
22}, 319. (11984).

F. Cattaneo, Astrophys. J. {\bf 434}, 200 (1994).

S.K. Chakrabarti, Mon. Not. R.A.S. {\bf 252}, 246 (1991).

B. Cheng, A.V. Olinto, D.N. Schramm and J.W. Truran, Preprint
Los Alamos National Laboratory (1997).

B. Cheng and A. Olinto, Phys. Rev. D {\bf 50}, 2421 (1994).

B. Cheng, D.N. Schramm and J.W. Truran, Phys. Rev. D {\bf 49},
5006 (1994).

M. Chiba and H. Lesch, Astron. Astrophys. {\bf 284},731 (1994).

P. Coles, Comments Astrophys. {\bf 16}, 45 (1992).

R.A. Daly and A. Loeb,  Astrophys. J. {\bf 364}, 451. (1990).

S. Davidson, Phys. Lett. B {\bf 380},253 (1996).

A.C. Davis and K. Dimopoulos, cern-th/95-175 (1995).

K. Dimopoulos and A.C. Davis, Phys. Lett. B {\bf 390}, 87 (1996).

A.D. Dolgov, Phys. Rev. D {\bf 48}, 2499 (1993).

A.D. Dolgov and J. Silk, Phys. Rev. D {\bf 47}, 3144 (1993).

J. Einasto, {\it Observational and Physical Cosmology}, Ed. by
F. Sanchez, M. Collados and R. Rebolo. Cambridge Univ. Press (1992).

J. Einasto, M. Einasto, S. Gottloeber, V. Mueller, V. Saar,
A.A. Starobinsky, E. Tago, D. Tucker, H. Andernach and P. Frsch, 
Nature {\bf 385}, 139 (1997).

K. Enqvist, in {\it Strong and electroweak
matter}. Hungary. astro-ph/9707300 (1997).

K. Enqvist and P. Olesen, Phys. Lett. B {\bf 319}, 178 (1993).

K. Enqvist and P. Olesen, Nordita preprint 94/6 (1994).

K. Enqvist, V. Semikoz, A. Shukurov and D. Shokoloff,
Phys. Rev. D {\bf 48}, 4557 (1993).

E. Florido and E. Battaner, Astron. Astrophys. in press (1997).

R.M. Gailis, N.E. Frankel and C.P. Dettman, Phys. Rev. D {\bf
52}, 6901 (1995).

W.D. Garretson, G.B. Field, S.M. Carrol, Phys. Rev. D {\bf 46},
5346 (1992).

M. Gasperini, M. Giovannini and G. Veneziano, cern-th/95-85
(1995).

M. Gasperini, M. Giovannini and G. Veneziano, cern-th/95-102
(1995b).

M. Gasperini and G. Veneziano, Astropart. Phys. {\bf 1}, 317
(1993).

M. Gasperini and G. Veneziano, Mod. Phys. Lett. A {\bf 8}, 3701
(1993).

D. Grasso and H.R. Rubistein, Astropart. Phys. {\bf 3}, 95
(1995).

D. Grasso and H.R. Rubinstein, Phys. Lett. B {\bf 379}, 73 (1996).

G. Greenstein, Nature {\bf 233}, 938 (1969).

E.H. Harrison, Mon. Not. R.A.S. {\bf 165}, 185 (1973).

J.C. Kemp, Pub. astron. Soc. Pacific {\bf 94}, 627 (1982).

P.J. Kernan, G.D. Starkman, T. Vachaspati, astro-ph/9509126
(1995).

T.W.B. Kibble and A. Vilenkin, Phys. Rev. D {\bf 52}, 1995
(1995).

K.T. Kim, P.P. Kronberg, G. Giovannini and T. Ventury, Nature
{\bf 341}, 720 (1989).

M. Hindmash and A. Everett, astro-ph/9708004 (1997).

C.J. Hogan, Phys. Rev. Lett. {\bf 51}, 1488 (1983).

A.M. Howard and R.M. Kulsrud, Astrophys. J. {\bf 483}, 648
(1996).

K. Jedamzik, V. Katalinic and A. Olinto, astro-ph/9606080 (1996).

E. Kim, A. Olinto and R. Rosner, Astrophys. J. {\bf468}, 28
(1996).

A. Kosowsky and A. Loeb, Astrophys. J. {\bf 469}, 1 (1997).

P.P. Kronberg, Rep. Prog. Phys. {\bf 57}, 325 (1994).

P.P. Kronberg, J.J. Perry and E.L.H. Zukowski, Astrphys.J. {\bf
387}, 528 (1992).

P.P. Kronberg and M. Simard-Normandin, Natura {\bf 263}, 653
(1976).

R.M. Kulsrud and S.W. Anderson, Astrophys. J. {\bf 396}, 606
(1992).

R.M. Kulsrud, S.C. Cowley, A.V. Gruzinov and R.N. Sudan,
Phys. Rep. {\bf 283},213 (1997). See also R. M. Kulsrud, R. Cen,
J. Ostriker and D. Ryn, Astrophys. J. {\bf 480}, 481 (1997).

K.M. Lanzetta, A.M. Wolfe and D.A. Turnshek, Astrophys. J. {\bf
440}, 435 (1995).

S. Lee, A. Olinto and G. Siegl, Astrophys. J. Lett. {\bf 455}, L21 
(1995).

D. Lemoine and M. Lemoine, Phys. Rev. D {\bf 52}, 1995 (1995).

H. Lesch and G. Birk, Phys. of Plasmas {\bf 5}, 2773 (1998).

H. Lesch and M. Chiba, Astron. Astrophys. {\bf 297}, 305 (1995).

H. Lesch and M. Chiba, Fundamentals of Cosmic Phys. {\bf 18},273
(1997).

H. Lesch, R. Crusius, R. Schlickeiser and R. Wielebinski,
Astron. Astrophys. {\bf 217}, 99 (1989). J.D. Huba and J.A. Fedder,
Phys. Fluids B {\bf 5}, 3779 (1993).

J.C.R. Magueijo, Phys. Rev. D {\bf 49},671 (1994).

J.J. Matese and R.F. O'Connell, Astrophys. J. {\bf 160}, 451
(1970).

C.J.A.P. Martins and E.P.S. Shellard, astro-ph/9706287 (1997).

T. Matsuda, H. Sato and H. Takeda, Pub. astr. Soc. Japan, {\bf
23}, 1 (1971).

I.N. Mishustin and A.A. Ruzmaikin, Sov. Phys. JETP {\bf 34}, 233
(1973).

P. Olesen, in {\it Nato advanced research workshop on ``Theretical
Physics''}, Zakopane. Poland (1997).

P.J.E. Peebles, {\it The large scale structure of the Universe},
Princeton Univ. Press. Princeton (1993).

J.J. Perry, A.M. Watson and P.P. Kronberg, Astrophys.J. {\bf
406}, 407 (1993).

J.H. Piddington, {\it Cosmic Electrodynamics}. Wiley
Interscience. New York (1969).

J.H. Piddington, {\it Cosmic Electrodynamics}. Wiley
Interscience. New York (1970).

R. Plaga, Nature, {\bf 374}, 430 (1995).

R. Pudritz, Mon. Not. R.A.S. {\bf 195}, 881 (1981).

R. Pudritz and J. Silk, Astrophys. J. {\bf 342}, 650 (1989).

J. Quashnock, A. Loeb and D.N. Spergel, Astrophys. J. Lett. {\bf 344},
L49 (1989).

B. Ratra, Astrophys. J. Lett. {\bf 391}, L1 (1992).

M. Rees, Q. Jl. R. astr. Soc. {\bf 28}, 197 (1987).

M.J. Rees and M. Reinhardt, Astron. Astrophys. {\bf 19}, 104
(1972).

A.A. Ruzmaikin, A.M. Shukurov and D.D. Sokoloff, {\it Magnetic
Fields of Galaxies}. Kluver, Dordrecht (1988).

S.L. Shapiro and I. Wasserman, Nature, {\bf 289}, 657 (1981).

S.A. Shectman, S.D. Landy, A. Oemler, D.L. Tucker, H. Lin,
P. Kirshner and P.L. Schechter, Astrophys. J. {\bf 470}, 172 (1996).

H. Sicotte, Mon. Not. R.A.S. {\bf 287}, 1 (1997).

G. Sigl, A. Olinto and K. Jedamzik, astro-ph/9610201 (1996).

J. Silk, Astrophys. J. {\bf 151}, 459 (1968).

T. Stanev, P.L. Biermann, J. Lloyd-Evans, J.P. Rachen and
A. Watson, Phys. Rev. Lett. {\bf 75}, 3056 (1995). T. Stanev,
Astrophys. J. {\bf 479}, 290 (1997). M. Lemoine, G. Sigl, A.V. Olinto
and D.N. Schramm, Astrophys. J. Lett. {\bf 486},  115 (1997). 

O. Tornqvist, hep-ph/9707513 (1997).

R.B. Tully, R. Scaramella, G. Vettolani and G. Zamorani,
Astrophys. J. {\bf 388}, 9 (1992).

M.S. Turner and L.M. Widrow, Phys. Rev. D {\bf 37}, 2743 (1988).

T. Vachaspati, Phys. Lett B {\bf 265}, 258 (1991).

T. Vachaspati and A. Vilenkin, Phys. Rev. Lett. {\bf 67}, 1057
(1991).

S.I. Vainshtein and F. Cattaneo, Astrophys.J. {\bf 393}, 165
(1992).

S.I. Vainshtein, E.N. Parker and R. Rosner, Astrophys. J. {\bf
404}, 773 (1993).

J.P. Vallee, Astrophys. Lett {\bf 23}, 87 (1983).

G. Veneziano, Phys. Lett. B {\bf 265}, 287 (1991).

I. Wasserman, Astrophys. J. {\bf 224}, 337 (1978).

S. Weinberg, {\it Gravitation and Cosmology} John Wiley 
Sons. New York (1972).

A.M. Wolfe, K.M. Lanzetta and A.L. Oren, Astrophys. J. {\bf 404},
480 (1992).

Ya B. Zeldovich, Sov. Phys. JETP {\bf 48}, 986 (1965).

Ya B. Zeldovich and I.D. Novikov, {\it The Estructure and Evolution
of the Universe}  Univ. Press, Chicago (1975).

A. Zurita and E. Battaner, Astron. Astrophys. {\bf 322}, 86
(1997).

E.G. Zweibel, Astrophys. J. {\bf 329}, L1 (1988).

B.G. Zweibel and C. Heiles, Nature {\bf 385}, 131 (1997).

\end{document}